\documentclass[preprint,showpacs,preprintnumbers,amsmath,ammsymb]{revtex4}

\draft

\begin{document}

\title{
Electronic structures and Raman features of a carbon nanobud 
}
\author{
         $H.Y. He^1$  and $B.C. Pan^{1,2}$
} 
\affiliation{
  1 Department of Physics,
  2 Hefei National Laboratory for Physical Sciences at Microscale,  \\
   University of Science and Technology of China,\\
 Hefei, Anhui 230026, People's Republic of China \\
}


\begin{abstract}
By employing the first-principles calculations, we investigate electronic properties of a novel carbon nanostructure called 
a carbon nanobud, in which a $C_{60}$ molecule covalently attaches or embeds in an armchair carbon nanotube. 
We find that the carbon nanobud exhibits either semiconducting or metallic behavior, depending on the size of 
the nanotube, as well as the combination mode. Moreover, with respect to the case of the corresponding pristine nanotubes, 
some new electronic states appear at 0.3-0.8 eV above 
the Fermi level for the carbon nanobuds with the attaching mode, which agrees well with the experimental reports.  
In addition, the vibrational properties of the carbon nanobuds are explored. 
The characteristic Raman active modes for both $C_{60}$ and the corresponding pristine nanotube present in Raman spectra 
of the carbon nanobuds with attaching modes, consistent with the observations of a recent experiment. 
In contrast, such situation does not appear for the case of the carbon nanobud with the embedding mode. 
This indicates that the synthesized carbon nanobuds are probably of the attaching configuration rather than 
the embedding configuration.
 
\end{abstract}


\maketitle

\section {INTRODUCTION}

Since the discovery of fullerenes and carbon nanotubes(CNTs), the composites of the low-dimensional carbon materials 
have attracted a great research interest, due to their unique physical 
and chemical properties.\cite{Treacy,Lieber2,Heer,Hebard} Recently, a novel hybrid carbon nanostructure named as 
carbon nanobud (CNB), in which a fullerene covalently bonded to the outer wall of a carbon nanotube, 
was synthesized in experiment. \cite{Nasibulin,Nasibulin-2} 
Different sizes of the fullerenes were probed in the CNBs, 
among which the component of $C_{60}$ was the largest. With respect to the cases of the individual fullerenes and carbon 
nanotubes, the measured scanning tunneling spectroscopy (STS) of the CNBs showed new peaks at about 0.3 and 0.8 eV 
above the Fermi level; Meanwhile, the observed adsorption spectra exhibited different 
features in the low energy range. These different features are essentially attributed to the covalent combination of 
the fullerenes with the carbon nanotubes. In other words, the combined fullerenes in the carbon nanotubes modulate the 
electronic structures of the systems significantly. Such modulation of the electronic structures is responsible for the 
observed new properties, such as high current density and a high cold electron field 
emission efficiency. As a result, it was suggested that CNBs are promising materials in 
field emission and nano-device in the future.

Physically, the observed advantageous properties of CNBs tightly correlate with 
the structural patterns of the CNBs. Up to data, two kinds of
structural patterns of the CNBs were proposed: (1) a carbon fullerene molecule embeds in the outer wall of a carbon
nanotube, where the carbon nanotube is actually imperfect in its structure (some carbon atoms are removed at the 
connection region); (2) A perfect fullerene attaches on the outer wall of a CNT. For convenience, 
we call the first kind of structural pattern of CNBs as the embedding configuration, and the second one as the 
attaching configuration. Wang's group \cite{Wang} discussed the CNBs with the embedding modes. 
They found that the CNBs with zigzag carbon nanotubes exhibit semiconducting behaviors, 
no matter whether the zigzag nanotubes
are metallic or semiconducting; while those with armchair carbon nanotubes are metallic still. 
Later on, Wu and Zeng \cite{Zeng} calculated the CNBs with the attaching modes. Their calculations predicted that
the most stable configuration was the $C_{60}$ molecule attaching on the outer wall of (5,5) or (10,0) nanotube via two 
C-C bonds tilted against the tube axis. Their calculated electronic structures showed that 
all the CNBs characterize semiconducting features regardless of the original CNTs being metallic or semiconducting.
It is noted that two theoretical groups handled the hybrid carbon systems using different structural patterns. Thus their
results can not be comparable with each other. Consequently, it is still difficult for us to judge which structural 
pattern is responsible for the observed CNBs in experiments.

Usually, to determine the favored structural features of CNBs obtained in experiment, it is necessary to calculate some properties 
of a CNB with each proposed atomic structural model, then to make comparison with the related properties observed 
in experiments. As mentioned above, the STS of the hybrid carbon materials provided new electronic states, 
which is available for our comparison. 
On the other hand, through measuring the Raman spectra of a CNB, Tian et al. observed the characteristic Raman features 
of both CNTs and fullerenes simultaneously. \cite{Tian} This is also a very important reference for us to 
explore the structural features of the CNBs. In this work, taking a CNB consisting of a (8,8) CNT 
and $C_{60}$ as an example, 
we calculate the electronic and vibrational properties of the CNBs with different structural patterns. We find 
that the electronic structures and Raman features of the CNBs with the attaching mode are 
consistent well with the experimental observations, which indicates that the synthesized hybrid carbon materials 
most probably characterize the attaching rather than the embedding features in structures.

\section {COMPUTATIONAL METHOD}

Our calculations are performed by the SIESTA program, \cite{Siesta} 
in which the norm-conserving pseudopotential and the Perdew-Burke-Ernzerhof 
generalized gradient approximation (GGA-PBE) \cite{Perdew} are taken into account.
A double- $\zeta$ basis set\cite{Soler} is used for $C$ atoms. 
For each concerned system, a periodic boundary condition along the carbon nanotube axis is applied. 
The supercell length along the tube axis is set to be 19.68 $\AA$, so that the interaction between 
the attached $C_{60}$ molecule and its images can be neglected. 
The geometries of the concerned systems are allowed to be fully optimized, 
with the residual force acting on each atom being less than 0.02 eV/$\AA$.
The Brillouin zone is sampled with $1\times1\times18$ according to the Monkhorst-Pack scheme, \cite{Monkhorst} 
which is tested to be enough in our calculations. 
In our phonon calculation, the frozen phonon approximation \cite{Yin} is employed, where 
the atoms in a system are displaced one by one from their
equilibrium positions with an amplitude of 0.04 Bohr. The calculation details refer to the previous work. \cite{He}

\subsection {STABILITY AND ELECTRONIC STRUCTURES}

Four kinds of configurations for the CNBs (shown in Fig.~\ref{Fig.1}) are considered, which are described as below: 
(1) two carbon atoms in a C-C bond between two hexagonal faces in $C_{60}$ bond with the 
atoms in a C-C bond $perpendicular$ to the tube axis in a (8,8) CNT; (2) two carbon atoms in a C-C 
bond between two hexagonal faces in $C_{60}$ bond with the atoms in a C-C bond $tilted$ to the tube axis in a 
(8,8) CNT; (3) a $C_{60}$ molecule connects with a (8,8) CNT via six C-C bonds in a hexagon ring; 
and (4) a $C_{60}$ molecule embeds in a (8,8) CNT, where six carbon atoms in the CNT at the connection
region are removed and the $C_{60}$ molecule just locates at this defective site of the (8,8) CNT. 
For convenience, we term these configurations as type-I, type-II, type-III and type-IV respectively. Of these 
configurations, type-I, type-II and type-III are sorted to the attaching mode, and type-IV to 
the embedding mode. 
After full relaxations, the formation energies of these configurations are evaluated to be 1.76, 1.27,
6.05 and 6.26 eV respectively. Here the formation energy $E_f$ is defined as $E_f=E_{CNB}-E_{CNT}-E_{C60}$,
where $E_{CNB}$ is the total energy of the CNB, $E_{C60}$ is the energy of an isolated $C_{60}$, and
$E_{CNT}$ is the energy of the corresponding CNT. Here, $E_{CNT}$ is a free-standing CNT for type-I,type-II 
and type-III configurations. For type-IV configuration, due to imperfect structure of the CNT, $E_{CNT}$ 
is evaluated by $E_{CNT}=N_{def}\times E_0$, where $N_{def}$ is the number of the carbon atoms in the 
defected CNT, and $E_0$ is the energy per atom of a perfect free-standing CNT. 
These calculated energies indicate that the type-I and 
type-II configurations are more stable than the others, of which the type-II configuration is more favorable.

To go further, we calculate the band structures and the density of states (DOS) of the concerned CNBs, 
as shown in Fig.~\ref{Fig.2}, where the calculated band structures and DOS of the pristine (8,8) CNT are 
also plotted as a reference. Clearly, the CNB with type-II configuration has a narrow band gap of about 0.12 eV, 
characterizing a semiconducting feature, which is similar to the previous report. \cite{Zeng} 
Differently, the CNB with either type-I or type-III configuration remains metallic behavior. 
Meanwhile, the CNB with the embedding configuration, type-IV, is semiconducting, with a band gap of about 0.18 eV. 
Basically, for the CNB with the embedding configuration, the semiconducting behaviors of the system 
is attributed to large alteration of the structures of both $C_{60}$ and the (8,8) CNT. 
While for the CNBs with the attaching configuration, it is not straightforward to 
make sense about configuration-dependent conductivity.
To understand such configuration-dependent conductivity, 
we set the spacing between $C_{60}$ and the wall of the CNT to be different values, followed by calculations of the 
electronic structure for each case. 
Our calculations show that for the case of type-II, when the distance (D) between $C_{60}$ and the (8,8) CNT 
is larger than 3.0 \AA, the system exhibits metallic behaviors;
Otherwise, the CNB displays semiconducting behaviors. In contrast, for the case of either type-I or type-III configuration, 
the metallic feature does not alter with variation of the distance D.
Careful examination of the local atomic structures around the connection sites reveals that 
there still exist two mirror symmetries (parallel and perpendicular to the tube axis) in both type-I and type-III 
configurations.
However, for type-II configuration, the symmetry of the initial CNT is broken at all by attaching $C_{60}$. 
Consequently, the CNB with type-II configuration exhibits a semiconducting behavior. 

In addition, by comparing the total density of states (black solid lines in DOS of Fig.~\ref{Fig.2} (b-e))
of the CNBs with that of the pristine (8,8) tube (Fig.~\ref{Fig.2} (a)), one can find that two apparent peaks marked with 
$A$ and $B$ appear at about 0.3-0.8 eV above the Fermi level for the cases of type-I, type-II and type-III. 
However, these peaks can not be observed in the case of type-IV, except for a peak marked with $A$ positions at 0.1 eV above 
the Fermi level. In comparison with the literature, \cite{Nasibulin}
the features in DOS around the Fermi level for the first three configurations seemingly match the observations in experiment,
whereas the configuration of type-IV does not yet. Accordingly, we suggest that the observed CNBs might not have the
configuration of type-IV.
For the CNBs with type-I, type-II or type-III configuration, further local density of states analysis reveals 
that the peaks of $A$ and $B$ are contributed from $C_{60}$, as shown in Fig.~\ref{Fig.2} (b-d). 
It is worth to note that these two peaks dose not appear in the DOS of an isolated $C_{60}$. In contrast, 
for type-IV configuration, the peak at 0.1 eV above Fermi level are from the contribution of 
both the (8,8) CNT and $C_{60}$ (Fig.~\ref{Fig.2} (e)). 

To address the electron behaviors near the Fermi energy, the charge density for the CNBs 
are given in Fig.~\ref{Fig.3}, where the highest occupied state (HOS) and the lowest 
unoccupied state (LUS) at $\Gamma$ point are plotted. For the CNBs with the attaching modes (type-I, type-II and type-III), 
the HOS is mainly contributed from the (8,8) CNT, and the LUS from the attached $C_{60}$. This is consistent with the 
DOS. In contrast, for the CNB with embedding mode (type-IV), both the CNT and $C_{60}$ make 
contribution to the HOS and LUS, which indicates 
that stronger coupling exists between CNT and $C_{60}$ for the embedding mode. This is clearly shown in the isosurface 
charge density.

In addition, we further explore the stability and the electronic properties of CNBs with different sizes of the CNTs. 
We find that for the CNBs combined with (5,5) and $C_{60}$, the formation energies are 0.75, 0.58, 4.88 and 4.08 eV 
for type-I, type-II, type-III and type-IV configurations respectively, which are comparable to 
the previous report. \cite{Zeng}. Moreover, the first three types of the CNBs all exhibit semiconductors, with the energy 
gap of 0.10, 0.05 and 0.20 eV, being also consistent with the results of Zeng's group. Interestingly, the CNBs with 
type-IV configuration exhibits metallic behavior.
Replacing the (5,5) CNT by a smaller (4,4) CNT, the CNBs are all of semiconductive properties. Moreover, the gap for 
type-III configuration is significantly broadened to be 0.38 eV. 
From above, we find that the size of the contained CNT significantly influences the electronic properties of the CNBs. 
For the CNBs with the attaching modes, due to the curvature effect of the CNTs, the interaction between $C_{60}$ and the CNTs  
becomes stronger, resulting in large distortion in the local structure. 
Such distortion breaks the symmetries of the systems, leading to the transition from the metallic 
CNTs to the semiconducting CNBs. For the CNBs with embedding mode, 
size-preference seemly exists between $C_{60}$ and the CNTs.  
Therefore, we predict that for the CNBs consisting of an armchair CNT and $C_{60}$, the electronic structures 
not only have a close relationship with the sizes of the contained CNTs, but also 
tightly correlate with the combination modes. 


\subsection {VIBRATIONAL PROPERTIES}

As we know, vibration can be regarded as fingerprint of a system, which is used to identify the atomic 
structure of the system. Thus we turn to study the vibrational properties of the typical CNBs combined with (8,8) and $C60$.
With using the frozen phonon approach, we calculate the vibrational frequencies and 
the corresponding eigenmodes of the CNBs respectively.

For comparison, we firstly explore the vibrational frequencies of a pristine (8,8) CNT. The radial 
breathing mode (RBM) for the (8,8) tube is found to be at 255 $cm^{-1}$, 
a little larger than that from the previous report. \cite{Kurti} 
For the CNBs, the frequencies of the RBMs show a red shift of less than 
15 $cm^{-1}$ with respect to that of the free-standing (8,8) tube. 
In addition, some new vibrational modes emerge in the frequency region of 900-1300 $cm^{-1}$,
where an isolated $C_{60}$ and a pristine (8,8) CNT contribute few vibrational states.
Through analyzing the corresponding eigenmodes, we find that these new states are attributed 
to the stretching modes for some longer C-C bonds near the distorted connection area. 
Moreover, the high frequency range of each system extends by about 20 $cm^{-1}$, which are ascribed to
the stretching modes of some shorter C-C bonds induced by the combined $C_{60}$.

With using the empirical bond polarizability model, \cite{Guha,Saito1} we calculate the intensity of 
Raman active modes for these CNBs at the phonon temperature of 300K. In our calculations, the laser with excitation
wavelength of 514.5 nm is used.
The obtained Raman spectra for the CNBs are shown in Fig.~\ref{Fig.4}, in which the Raman spectra 
for an isolated $C_{60}$ and a pristine (8,8) tube are plotted for comparison. 

In Fig.~\ref{Fig.4} (d), the typical Raman active modes of R band and G band for the (8,8) 
CNT are predicted to be at 255 and 1673 $cm^{-1}$, which are very close to those of the previous report. \cite{Rao} 
Meanwhile, a pronounced peak at 1627 $cm^{-1}$ corresponding to the stretching modes of the C=C bonds 
and a weaker peak at 485 $cm^{-1}$ corresponding to 
the pentagon breathing mode are also found in the Raman spectrum of the isolated $C_{60}$ molecule, 
which is in agreement with the previous reports. \cite{Bethune} 
This indicates that our Raman analysis is reliable. 

Interestingly, from the Fig.~\ref{Fig.4} (a-c), one can see clearly that the characteristic Raman features 
for the isolated (8,8) tube and $C_{60}$ (Fig.~\ref{Fig.4} (d)) as mentioned above simultaneously appear 
in the Raman spectra of the CNBs with type-I, type-II and type-III configurations, 
only with frequency shifts within 10 $cm^{-1}$.
In contrast, for the case of type-IV configuration, these characteristic Raman features for both $C_{60}$ and 
the (8,8) CNT disappear or become weak to some extent. Instead, more Raman active modes are introduced by defective 
structure. 
Such difference of Raman spectra is closely related with the structural features of the CNBs. For 
CNBs with the attaching modes, both $C_{60}$ and the (8,8) CNT have the perfect structure,
in which covalent combination of $C_{60}$ and the CNT has a weak influence on Raman features. 
While for the latter case, either $C_{60}$ or the (8,8) CNT has imperfect structure, which influences the Raman 
features of the system significantly. 
Recall that the typical modes of $C_{60}$ and an isolated CNT were identified in the measured Raman 
spectra of the CNB, \cite{Tian} thus our calculations suggest that the type-IV configuration may be ruled out 
for the produced CNBs.

\section {Conclusion}

In summary, based on the density functional theory calculations, we explore electronic properties of the CNBs 
consisting of $C_{60}$ and an armchair CNT with four kinds of configurations. 
We find that the CNBs exhibit either metallic or semiconducting conductivity, which is dependent on 
the combination mode between $C_{60}$ and the CNT, as well as the size of the contained CNT.
Furthermore, our calculations show that the typical Raman features of the isolated $C_{60}$ 
and the pristine carbon nanotube still appear in the Raman spectra of the CNB with type-I, type-II 
or type-III configuration, but do not appear in that of the CNB with type-IV.
As compared with the STS and Raman spectra in the experiments, our theoretical calculations predict that 
the observed CNBs in experiments were most probably with the attaching configuration rather 
than the embedding configuration.

\section *{Acknowledgements}   

 This work was partially supported by the National Basic Research Program of China, 
Grant No. 2006CB922000, 2009CB939901, and by the China Postdoctoral Science Foundation with 
code number of 20090450813.

\newpage

\begin{figure}
\caption{(Color online) The relaxed configurations of the carbon nanobuds with (a) type-I, (b) type-II, 
(c) type-III or (d) type-IV mode.}\label{Fig.1}
\end{figure}
                                                                                                                             
\begin{figure}
\caption{(Color online)
The bands structures (left) and DOS (right) for (a) the pristine (8,8) tube, the carbon nanobuds with (b) type-I, 
(c) type-II, (d) type-III or (e) type-IV configuration. On the right of each figure, the solid lines 
stand for total density of states, and the dash lines and the dot-dash lines indicate local density of states 
for $C_{60}$ and the (8,8) CNT respectively. The dot lines in each figure refer to Fermi level, which are shifted 
to zero.}\label{Fig.2}
\end{figure}
   
\begin{figure}
\caption{(Color online)
The isosurface charge density of the highest occupied state (HOS) and the lowest unoccupied state (LUS) for the 
CNBs combined with $C60$ and (8,8) with (a) type-I, (b) type-II, (c) type-III or (d) type-IV configuration.
 The isosurface value is $\pm$ 0.04 eV/$\AA^{3}$ (distinguished by grey and blue surfaces). 
}\label{Fig.3}
\end{figure}

\begin{figure}
\caption{(Color online)
The Raman intensity for the carbon nanobuds with (a) type-I, (b) type-II,
(c) type-III, (d) type-IV and for (e) an isolated $C_{60}$ and (8,8) CNT. 
The solid lines and the dash lines in (d) stand for 
the Raman spectra of $C_{60}$ and the (8,8) tube respectively. }\label{Fig.4}
\end{figure}

\widetext

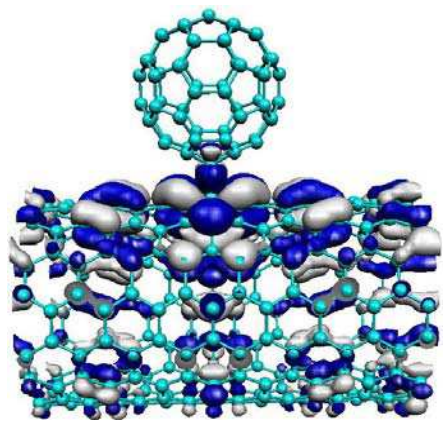
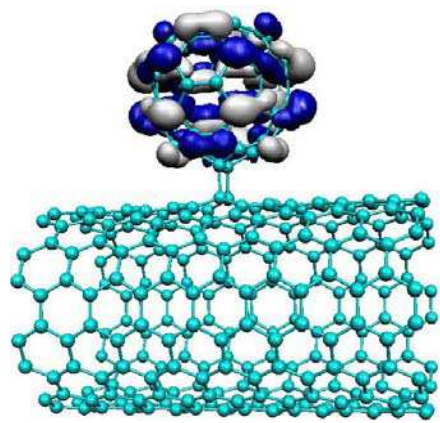
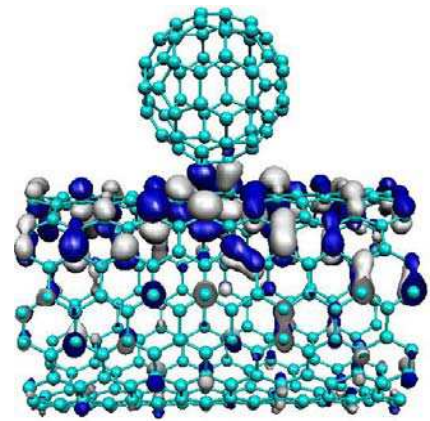
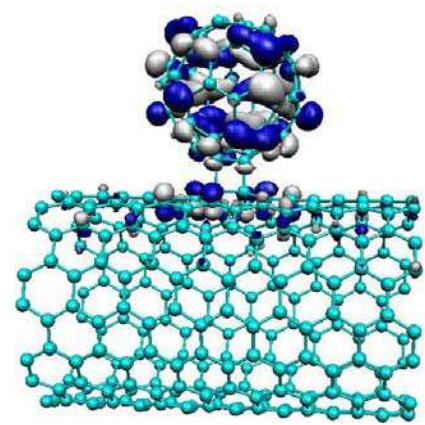
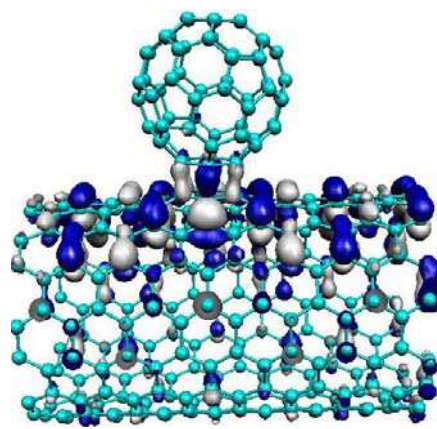
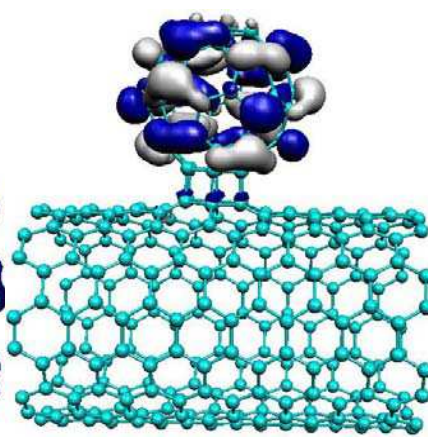
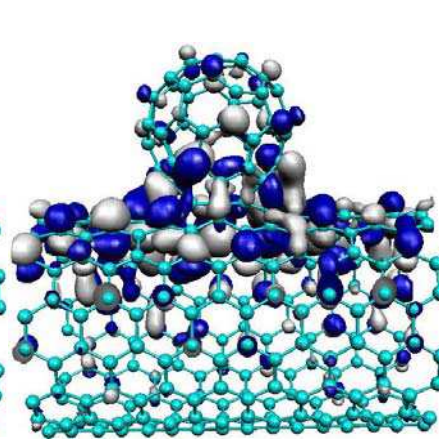
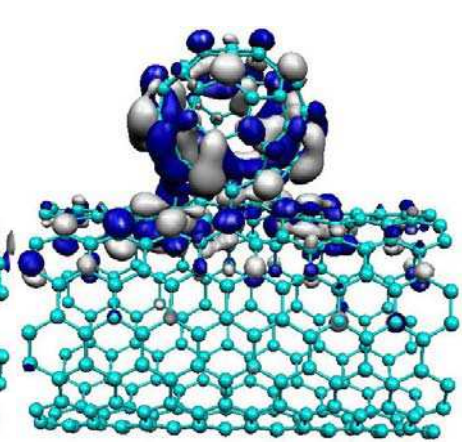
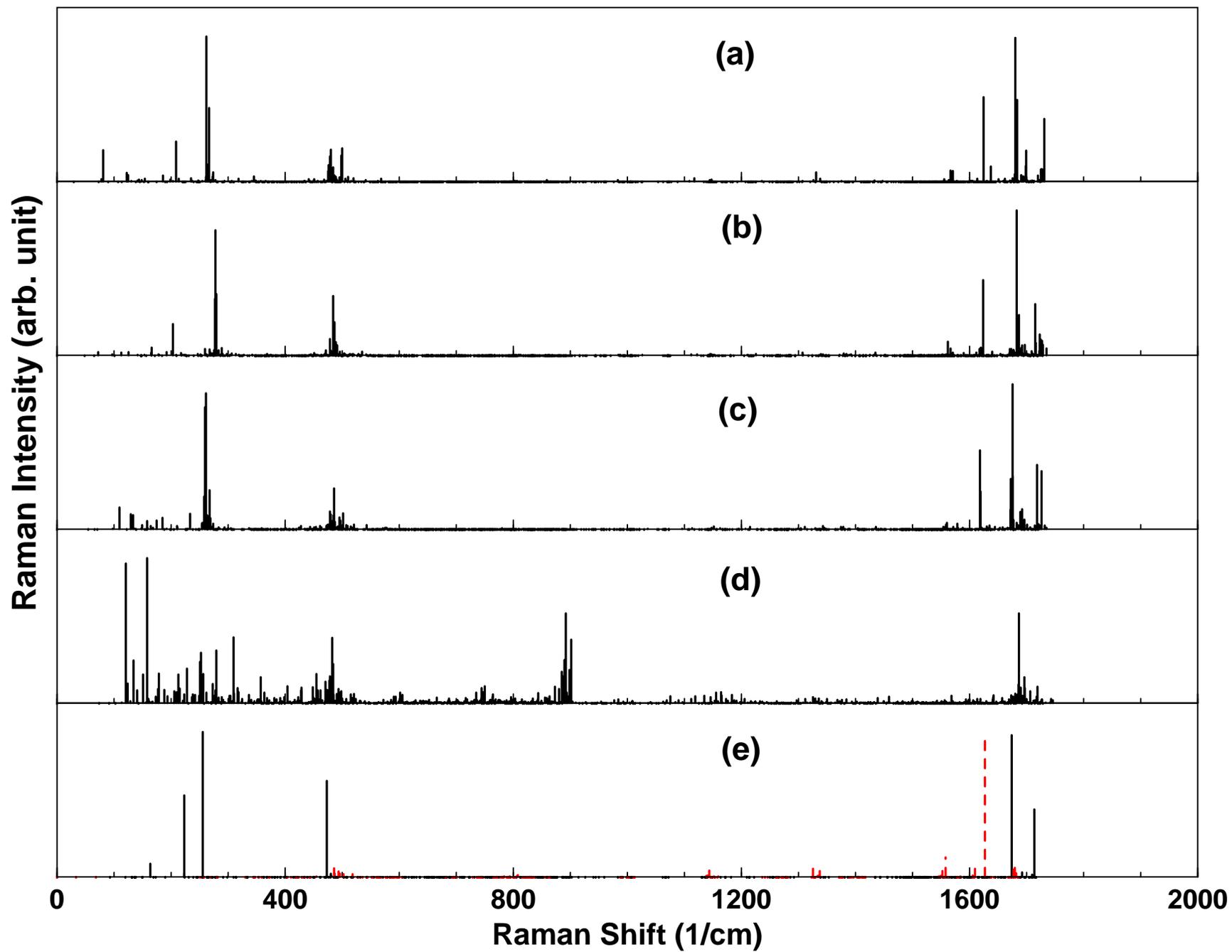
\end{document}